# Hybrid Decision Algorithm for Access Selection in Multi-Operator Networks

Soha Farhat[1,2], Abed Ellatif Samhat[1], Samer Lahoud[2], Bernard Cousin[2]
[1]Lebanese University, Ecole Doctorale des Sciences et Technologies, Hadath, Lebanon
[2]University of Rennes I – IRISA, France

*Abstract*— In this paper, we propose a hybrid decision algorithm for the selection of the access in multi-operator networks environment, where competing operators share their radio access networks to meet traffic and data rate demands. The proposed algorithm guarantees the user satisfaction and a global gain for all cooperating operators. Simulation results prove the efficiency of the proposed scheme and show that the cooperation between operators achieves benefits to both users and operators; user acceptance as well as the operator resource utilization and the operator revenue increase.

## I. Introduction

Wireless Networks are growing rapidly to support the heavy mobile broadband traffic, and to meet the ever-increasing user expectations. The operators feel the need to upgrade their networks in order to increase capacity, data rates and coverage, therefore satisfying their subscribers with maximum efficiency. To meet traffic and data rate demands, one approach is to deploy a heterogeneous network [1][2] with advanced traffic management. This cost-effective solution integrates multiple radio access technologies (RATs), such IEEE 802.11 WLANs, UMTS, LTE…, and assumes an efficient Radio Resource Management (RRM) jointly done among the different RATs. It creates a multi-RATs environment under the management of a single operator, and a joint RRM begins by optimizing users association to the different RATs in order to enhance resource utilization, users' satisfaction and overall network performance.

Another approach is to deploy a multi-operator sharing network [2], where competing operators share their radio access networks. This solution seems convenient, when there is insufficient revenue for the operator to deploy multiple networks. Moreover, it helps to avoid the waste of radio resources, when traffic level is lower than planned, and to defeat QoS degradation, when the traffic is higher than expected. In such cooperative environment, when an operator is unable to insure satisfaction constraints to its user, he tries to give him access to the service through another network operator, thus avoiding his rejection. Therefore, operators' cooperation is unavoidable in order to achieve a joint resource management and consequently improve the global system performance and ameliorate operators' profits. The choice of the cooperating operator for the client transfer and the decision process are based on a selection problem, relying on different criterions and depending on specific user/operator considerations.

Most of the existing works have studied the access selection in the context of multi-RATs under a single operator, and many approaches were adopted to perform the decision [3-8][11]. To our knowledge, however, the multi-operator context is rarely considered. In this paper, we propose a hybrid decision method for RAT selection in a multi-operator context. The proposed selection algorithm is hybrid *i.e.*, it guaranties users' satisfaction, in terms of preference and QoS application requirements, and associates them in a way to improve operator's profits and network performance.

The rest of the paper is organized as follows: Section II presents some existing work related to RAT selection algorithms. Section III describes our hybrid decision algorithm. Simulation environment and results are presented in section IV. Section V briefly concludes the paper.

## II. Background and Related Works

Access Selection was widely studied in heterogeneous wireless networks with a single operator. Different approaches were adopted for decision making, and many algorithms were conceived in order to associate users to the best available RAT, at admission or during a vertical handover [3-8][11].

In order to select the best radio access network, a number of parameters must be considered. These parameters can be divided into four categories: Access network information, user preferences, terminal capabilities and service type. In a cost function based algorithm, these parameters are normalized, assigned a weight and then injected into a weighted sum to produce a selection score [3][8][11]. In [4], authors use fuzzy logic to deal with imprecise criteria and user preferences; data are first converted to crisp numbers and then classical Multiple Attribute Decision Making (MADM) methods as Simple Additive Weighting (SAW) and Technique for Order Preference by Similarity to Ideal Solution (TOPSIS), are applied. Another approach aims to prioritize the available RATs to decide the optimum one for mobile users. Such approach is used in [5]: Analytical Hierarchy Process (AHP) was adopted to arrange selection parameters in three hierarchical levels, in order to calculate the corresponding weighting factors. Then, Grey Relational Analysis (GRA) is applied to prioritize the networks for the selection decision. In [6], a performance comparison was made between Multiplicative Exponent Weighting (MEW), SAW, TOPSIS and GRA. Results showed similar performance to all traffic classes. However, higher bandwidth and lower delay are provided by GRA for interactive and background traffic classes. A network centric approach is adopted in [7], and a cost function is used to accommodate the maximum number of users in available RATs, while insuring load balancing. It consists of minimizing the costs of resource underutilization, demand rejection, thus maximizing the network operator profits. A Joint Radio Resource Management (JRRM) framework in a multi-operator environment is introduced in [9]. Authors extended their single operator approach to an operator cooperation scenario. They proposed a two-layer JRRM strategy to fully exploit the available radio resource and to improve operators' revenue. The proposed economic-driven JRRM is based on fuzzy neural methodology with different classes of input parameters: Technical inputs, Economic inputs and Operator policies. The satisfying RAT is selected referring to a Fuzzy Selected Decision (FSD) indicating the appropriateness of selecting each available RAT in front of the others. This work showed how an inter-operator agreement can bring more benefits in terms of network performance and operators' revenue. In [10], game theory is used for Dynamic Spectrum Access algorithm with cellular operators. Authors have defined a utility function, for the operators, considering user's bit rate, the blocking probability and the spectrum price. And they have presented a penalty function to control the blocking probability.

Therefore, we need a new algorithm for operator selection in a multi-operator environment, providing users the capability of being Always Best Connect (ABC) and guaranteeing operator's satisfaction, while enhancing the network performance. In our proposal, a user has more chance to get the service even when its home operator (which user has contract with) cannot satisfy his QoS requirements. In this case, the selection algorithm will direct him to the operator that best suit its demand while guaranteeing higher profit for his home operator. In fact, our proposal adds the profit by user exchange as a selection criterion, and it is assigned a specific weight $W_{Op}$ which can be varied to control selection and achieve more profits. Moreover, the proposed algorithm takes into consideration users' preferences, incomes and costs that may results from user's exchange, in order to optimize users association and guarantee profit improvement for all cooperating operators. Our hybrid approach for RAT selection in a multi-operator environment is presented in the following section.

### III. OPERATOR SELECTION ALGORITHM

Consider a set of mobile operators who decided to cooperate. After executing the selection algorithm, a user can be served in the network of the home operator, denoted by H-op, bound by a contract, or in the network of another operator, denoted by serving operator. Users are not aware of the transfer between operators.

*A. Decision Parameters*

In order to insure an ABC profile for users and guarantee operators' satisfaction, the selection decision must take into account a number of parameters collected from application requirements, the user and the available access networks.

The application requirements are specified based on QoS classes. We consider two classes: real time such as Conversational class and non-real time as the Interactive class [6]. The Conversational class groups services such as Voice over IP. This group of application is sensitive to jitter and delay. The Interactive class concerns the traffic of human interaction with remote equipments such as web browsing and server access. This scheme includes services sensitive to loss rate and round trip delay. Thus, four QoS parameters are considered as user application requirements [3]: the required Bandwidth $BW_{Req}$, the required Jitter $J_{Req}$, the required Delay $D_{Req}$, and the required Bit Error Rate $BER_{Req}$.

The second factor in the operator selection process is user preferences. In fact, user preferences are generally difficult to assess. The user may be ready to pay any price to receive service with best quality. Or he may prefer a cheap service regardless of the quality of the offered connection. In our work, user preferences are translated with a couple of weights [$W_{QoS}$ $W_p$] determining the degree of preference of the required QoS over the paid price per service $p$.

Furthermore, the wireless environment contains different operators as stated previously. And, without loss of generality, each operator manages a single RAT and is capable of offering service to all possible users. Each technology offers some QoS parameters such as: The available Bandwidth $BW_R$, delay specifications as the mean Jitter $J_M$ and the mean Delay $D_M$, and loss rate as the mean Bit Error Rate $BER_M$. In addition, each technology sets a service access price $sp$. Moreover, each operator adopts its own market strategy *i.e.*, an operator may consider user satisfaction as a top priority to prevent any churn risk, or he may ensure an acceptable QoS for his client while maximizing his profits. This will affect surely the selection decision for user transfer. In addition, the operators' cooperation could be done with a previous financial agreement determining the inter-operator service pricing, it means that when an operator transfers its client to another serving operator, the latter will charge him a service access cost $Cs$. This cost may be equal, lower or even higher than the service price $sp$. Therefore, an operator transferring his client must consider a serving operator with low service cost $Cs$, in order to ameliorate his profits. All the mentioned requirements and parameters will be quantified and injected in a cost function to achieve the selection decision.

*B. Decision Cost Function*

Since we intend to adopt a hybrid approach, the selection problem must fulfill two objectives: user's satisfaction and operator's satisfaction.

*1) User's Satisfaction:* We suppose that a user intends to connect for a single service. The admission request to his H-operator contains information about the service type and his preferences Fig.1. Once the H-operator cannot meet the application requirements of his client, because of lack of resource or an usatisfying QoS, the application request will be transferred to the available cooperating operators. The selection process associates the user to the best operator in a transparent way. The selected operator must offer satisfaying QoS specifications and must respect user preferences.

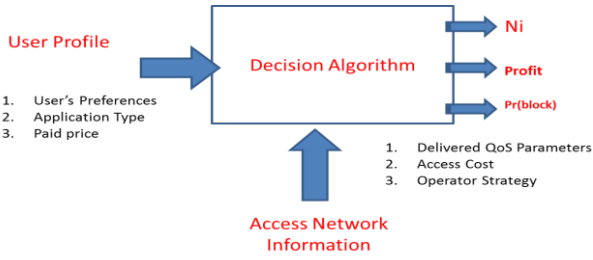

Fig. 1. Decision Parameters

However, choosing always the best network operator may penalize this network by an overload and the others by under-utilization. Therefore, we suggest choosing the operator delivering enough QoS to fit user's application requirements. To achieve this selection, we consider the Nearest Performance Handover introduced in [3] and used in the context of multi-RAT under single operator. It consists of defining a score for the ideal solution, then defining a score for every selection candidate, and finally, choosing the candidate with the closest score to the ideal one. In our approach, the ideal operator/network is the one delivering the QoS parameters required by the user's application. Hence, the ideal solution will be assigned the user score. This score is a weighted combinaition of the different required QoS parameters. We propose adding the paid price $p$ to this score in order to consider user preferences. Consequently, user's score $Su$ is computed as follows:

$$Su = W_{QoS} * S_{QoS} + W_p * p \quad (1)$$

Where, $W_{QoS}$ is the weight determining the degree of preference of the required QoS over the paid price, and $W_p$ is the weight determining the degree of preference of the paid price over the QoS. Moreover, $S_{QoS}$ is the user QoS score calculated as follows:

$$S_{QoS} = W_{JReq} * J_{Req} + W_{DReq} * D_{Req} + W_{BWReq} * BW_{Req} + W_{BERReq} * BER_{Req} \quad (2)$$

Where, $J_{Req}$, $D_{Req}$, $BW_{Req}$ and $BER_{Req}$ are the required jitter, delay, bandwidth and BER respectively, for user's application. These parameters are determined from the application QoS class, normalized and associated to their corresponding weights $W_{JReq}$, $W_{DReq}$, $W_{BWReq}$ and $W_{BERReq}$ respectively, before computing $Su$.

The selected operator that satisfies the user has a score with minimum distance to the ideal solution score. Every cooperating operator able to satisfy the user will be assigned a score $S_T$ computed as follows:

$$S_T = W_{QoS} * S_{TQoS} + W_p * sp \quad (3)$$

With

$$S_{TQoS} = W_{JM} * J_M + W_{DM} * D_M + W_{BWR} * BW_R + W_{BERM} * BER_M \quad (4)$$

Where $J_M$, $D_M$, $BW_R$ and $BER_M$ are the mean jitter, mean end-to-end delay, the remaining bandwidth and the mean loss rate measured on the candidate operator network, respectively. These parameters are normalized, assigned a weight, and then combined to form $S_T$.

Finally, the selected operator must satisfy the user by minimizing $|Su-S_T|$.

*2) Operator's Satisfaction:* An operator is satisfied when he gets better profits and improves his network performance. Thus, the preferred candidate to transfer his client (for service renting) will be an operator having a low $Cs$. In fact, a simplified cost analysis determines the income of the H-operator by the price paid by his client $p$ and the cost charged to him, while transferring his client, is the service cost $Cs$ established by the candidate serving operator. Consequently, the selected serving operator is the one capable of maximizing the profit ($p-Cs$).

Finally, considering both objectives, the best operator to choose is the one verifying the following condition:

$$Selected(Op_i) = \min_i (W_u * |Su-S_T| - W_{Op} * (p-Cs)) \quad (5)$$

Where, $W_u$ is the weight determining the degree of importance, for the H-operator, to satisfy the user. And $W_{Op}$ is the weight determining the degree of importance, for the H-operator, to improve his profits.

## IV. SIMULATION AND RESULTS

### A. Simulation Environment

For illustration, we consider three cooperating operators, $Op_1$, $Op_2$ and $Op_3$, each managing a single radio access network *UMTS*, *WLAN1* and *WLAN2*, respectively. The conditions of the networks are shown in Table I. In this study, we suppose that all RATs are capable of delivering a constant value for $J_M$, $D_M$ and $BER_M$. The normalization of the different parameters is done for each access network with respect to the service requirements. The normalized values of these parameters are presented in Table II. For the service price $sp$, we use the following values: 0.9, 0.1 and 0.2 unit/kByte [5] for $Op_1$, $Op2$ and $Op_3$ respectively.

After they arrive, mobiles are uniformly associated with a user profile, determining the service type, user preferences and the service price to pay $p$. We consider two possible service types: real-time and non real-time, and two preference vectors: [0.7 0.3] and [0.4 0.6], for high QoS preference and for high price preference, respectively. In addition, Conversational and Interactive QoS weights, corresponding to the bandwidth, the jitter, the delay and the loss rate are determined by applying AHP [5][6], and are given by the following vectors: [0.05, 0.45, 0.45, 0.05] and [0.16, 0.04, 0.16, 0.64], respectively.
Moreover, application requirements for real-time and non real-time services are determined in Table III, taking into account that bandwidth consumption of a service varies with the network technology.

TABLE I. UMTS AND WLAN NETWORK PARAMETERS [5]

| Network Technology | QoS Parameters | | | |
|---|---|---|---|---|
| | Bandwidth(Kb/s) | Jitter(ms) | Delay(ms) | BER(dB) |
| UMTS | 1700 | 6 | 19 | $10^{-3}$ |
| WLAN1 | 11000 | 10 | 30 | $10^{-5}$ |
| WLAN2 | 5500 | 10 | 45 | $10^{-5}$ |

TABLE II. NORMALIZED NETWORK PARAMETERS

| RAT | QoS Parameters[a] | | | |
|---|---|---|---|---|
| | BW | Jitter | Delay | BER |
| UMTS | 1 | 1 | 1 | 1 |
| WLAN1 | 1 | 1 | 1 | 1 |
| WLAN2 | 1 | 1 | 1 | 1 |

[a.] Normalization is done with respect to real-time service rquirements

TABLE III. APPLICATION REQUIREMENTS

| Service Type | QoS Parameters | | |
|---|---|---|---|
| | Jitter(ms) | Delay(ms) | BER(dB) |
| Real-Time | 10 | 100 | $10^{-3}$ |
| Non Real-Time | 20 | 150 | $10^{-5}$ |

### B. Simulation Setup

We consider the system formed by the three operators $Op_1, Op_2$ and $Op_3$. Users arrive to the system sequentially. We model the arrival and departure of users as a Poisson Process with mean arrival interval $1/\lambda$. We perform simulation for different values of $1/\lambda$= 5/2, 25/9, 10/3 and 5. Once connected, the user will stay in the system for a certain service time, assumed to follow an exponential distribution of mean $1/\mu$=4min. During this service time, the user will consume a constant bit rate depending on his service type and the access technology of the serving operator. Note that no scheduling is considered. And, at the end of the connection, the user will leave the system improving, consequently, the available bandwidth of the serving operator. The simulation is done for duration of 1200 sec and repeated for 20 experiments. We use Matlab to achieve this simulation.

### C. Simulation Results

For lack of space, we will present only simulation results for the system performance and profits amelioration, obtained for $W_u/W_{op}$=1 and $Cs=sp$, for all operators. The effect of user preferences and the financial strategy of the H-operator will be presented in a following work. Simulation results are discussed in this section. In all figures, the number of users represents the sum of arrivals in the system.

*1) Global performance:* We first illustrate simulation results for the global blocking probability, translating the global performance of the system. Fig. 2 compares the performance of the system, in term of blocking probability, before and after cooperation. Operators are assumed adopting the same strategy ($W_u/W_{op}$) and keeping the same price for service renting ($Cs=sp$). Simulation results show an excellent reduction in the blocking probability. When the operators cooperate, this probability does not reach 5%. One can see how much cooperation has helped operators to face blockings. These blockings, generally, are associated with a lack of resource, a bad QoS specifications or an overload situation. But, when operators cooperate, the blocking probability is reduced about 20%, showing that this cooperation is unavoidable and it brings benefits for the overall mobile system.

*2) Network performance:* The study of the blocking probability of each operator shows an important improvement, especially for the operators managing a limited capacity. Fig. 3 presents a comparison between the blocking probabilities, before and after cooperation, for every cooperating operator. One can see that, *Op1* is taking the largest benefit from this cooperation. His blocking probability is reduced by 48%, thus raising the number of admitted users. *Op1* could face overload situations by transferring his clients to *Op2* and *Op3*. In addition, *Op3* has limited his blocking percentage below 2% after cooperation. Moreover, *Op2* has benefitted slightly of this cooperation; this operator already had a low blocking rate even without cooperation. In fact, *Op2* has the best capacity among the cooperating operators in the system.
In addition, comparing the blocking probability for *Op2* and *Op3* shows the same performance for a number of arrivals below 360 users, but for higher number of arrivals, *Op3* denotes higher blocking rates. This is due not simply to the capacity difference, but also to the fact that *Op3* is more acting as a serving operator than *Op2*. An additional analysis of the served users by *Op2* and *Op3*, for high arrival rates, showed that up to 40% of the users served by *Op3*, are guests coming mostly from *Op1*, but *Op2* guests do exceed 18% of the served users. Beside, when *Op1* intends to transfer his clients, 79% of them go towards *Op3* and the rest to *Op2*. In addition, the majority of the transferred clients, from *Op1* to *Op3*, seeks an interactive service that consumes well in a WLAN RAT, thus raising the probabilty of blocking in *Op3* access network.

*3) Operators' profit amelioration:* The capacity gain achieved through cooperation induced a profit gain. New incomes are available from guests and transferred clients. Fig. 4 shows profit improvements for *Op1*, *Op2* and *Op3*. The important increase of the users' acceptance, after cooperation, brought more incomes for *Op1*; clients transferred, from *Op1* to another serving operators, instead of being blocked and guests served by *Op1* participated all in the increase of *Op1* incomes. We can notice the remarkable amelioration of *Op1* profits. *Op3* also benefits from profit amelioration. In fact, up to 40% of the incomes of *Op3*, as shown in Fig. 5 are from exhanged clients and served guests. This income has risen after cooperation. In addition to the increase of users' acceptance, *Op3* exchanges 90% of his transferred clients with *Op2*, having lower cost *Cs* than *Op1*. Another study of *Op3* transferred clients shows that, 95% of the clients transferred to *Op2* seek non real-time services and all transferred clients to *Op1* seek real time services Table IV. Consequently, it reduces the rate consumption, thus the service cost charged to *Op3*, next to this exhange. We can see clearly from Fig. 5 that the service cost charged to *Op3* is lower than the additionnal incomes from tranferred users. Again, *Op2* does not take much benefits from this cooperation, little profit improvement is noticed. If we remark the amount of users exchanged (transferred) from *Op2* in Fig. 5, we can see that it is relatively low. Moreover, a study of the direction of this transfer has revealed that the majority of the transferred clients are designated to *Op1*. The latter has high service cost; 9 times the *Cs* of *Op2*, thus increasing the cost service charged to *Op2* until it exceeds the incomes gained from users' exchange, Fig. 5. It is obvious, in this scenario, that *Op2* is not benefiting directly from this cooperation. But, since *Op2* has served guest users, we can say that he has benefitted by avoiding resource underutilization. *Op2* can achieve more profits by controlling his service cost *Cs* and set a higher price for service renting.

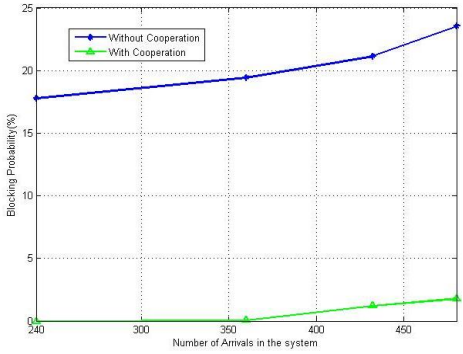

Fig. 2. Global Blocking Probability Comparison

*4) Sensitivity Analysis:* Table IV shows the direction of application (client) exchange. For *Op1*, all clients seeking a real-time service are migrated to *Op3*. The latter can guarantee a good QoS for this type of service. Moreover, when the transferred clients of *Op1* are seeking a non real-time, they are distributed on *Op2* and *Op3* networks. A deeper analysis of this transfer showed that clients are moved to *Op3* instead of *Op2*, only when *Op3* is the unique choice for exchange. For *Op2*, the majority of real-time services are transferred to UMTS technology delivering the best jitter, but capacity limitations of *Op1* caused that a portion of those services is moved to *Op3*. In case of non real-time services, all clients are moved to *Op3* delivering better QoS specifications, espacially that UMTS technology offers high BER for this kind of service. For *Op3*, the real-time services are distributed on *Op1* and *Op2* networks, since *Op1* has limited capacity. But all non real-time services are moved to *Op2*, offering better BER. We can say that, the selection algorithm could assign the best operator/network, and respect the QoS demand without penalizing the best operator/network by an overload.

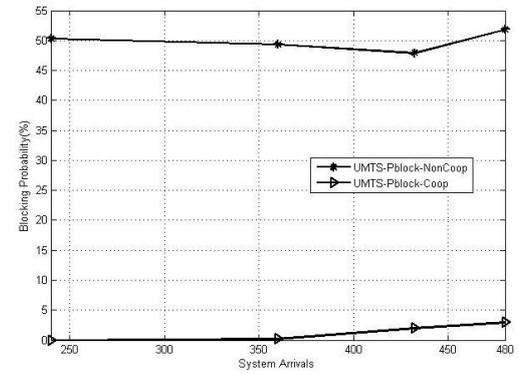
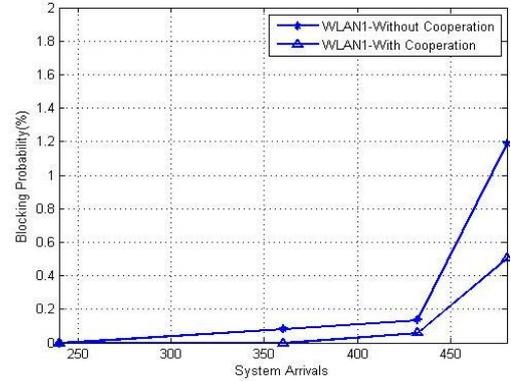
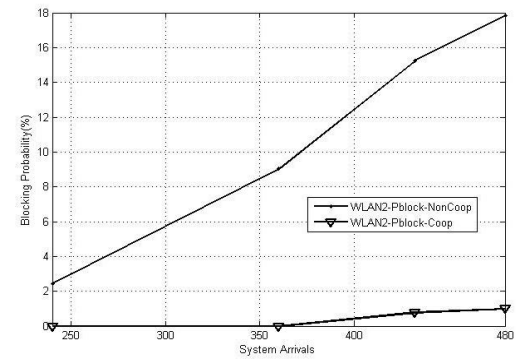

Fig. 3. Operators' Blocking Percentage Comparison

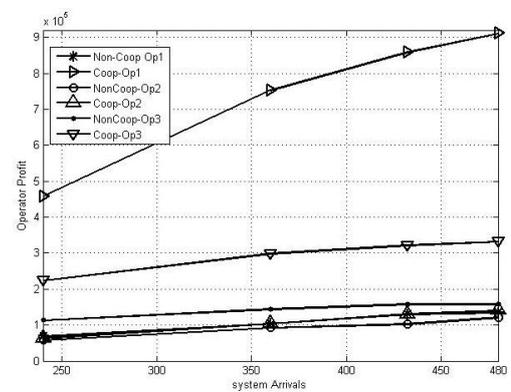

Fig. 4. Operators' Profits before and after cooperation

V. CONCLUSION

In this paper, a hybrid decision algorithm for the selection of the access in a multi-operator network has been presented. A cost function has been used. It combined the performance information given by the different wireless network and the requirements of the mobile user's application, added to the resulting profit of the user exchange. Moreover, it considered user preferences and operator's strategy, in order to guarantee the ABC user profile and a global gain for all cooperating operators. Simulation results proved the efficiency of the proposed scheme in terms of user and operator satisfaction, load balancing and network performance enhancement. Future work will investigate the effect of the weights related to operator strategy [$W_u$  $W_{Op}$], and the pricing policy, to show how the operator can achieve more revenues and ameliorate its profits, while cooperating.

TABLE IV. DIRECTION OF APPLICATION EXCHANGE

| Application Type From \ To | Real-Time | | | Non Real-Time | | |
|---|---|---|---|---|---|---|
| | *Op1* | *Op2* | *Op3* | *Op1* | *Op2* | *Op3* |
| *Op1* | - | 0 | 100% | - | 20% | 80% |
| *Op2* | 90% | - | 10% | 0 | - | 100% |
| *Op3* | 10% | 90% | - | 0 | 100% | - |

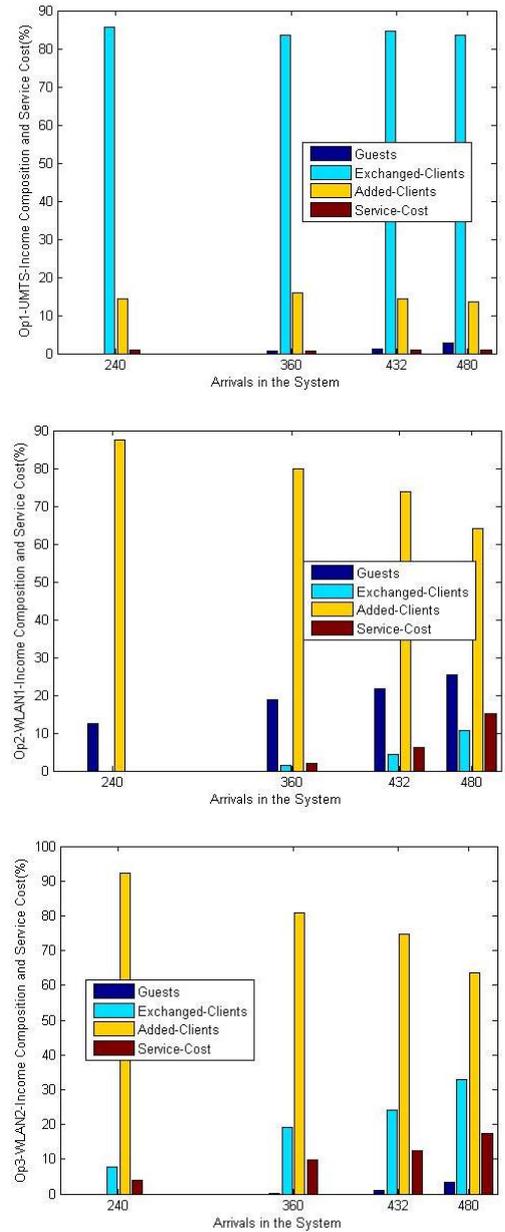

Fig. 5. Operators' Income Composition and Service Cost after Cooperation